# Note: An avalanche transistor-based nanosecond pulse generator with 25 MHz repetition rate

Nikolai Beev [a], Jonas Keller [b], Tanja E. Mehlstäubler

*QUEST Institute for Experimental Quantum Metrology, Physikalisch-Technische Bundesanstalt. Bundesallee 100, 38116 Braunschweig, Germany*

(Dated: December 11, 2017)

We have developed an avalanche transistor-based pulse generator for driving the photocathode of an image intensifier, which comprises a mainly capacitive load on the order of 100 pF. The circuit produces flat-top pulses with rise time of 2 ns, FWHM of 10 ns and amplitude of tens of V at a high repetition rate in the range of tens of MHz. The generator is built of identical avalanche transistor sections connected in parallel and triggered in a sequence, synchronized to a reference rf signal. The described circuit and mode of operation overcome the power dissipation limit of avalanche transistor generators and enable a significant increase of pulse repetition rates. Our approach is naturally suited for synchronized imaging applications at low light levels.

Generators of nanosecond-scale pulses of voltage or current are used to drive electro-optical devices such as diode lasers [1], Pockels cells [2], electro-optical modulators [3], and image intensifiers [4]. Bipolar transistors operated in avalanche breakdown mode are commonly used to construct such generators [1] [4] [5] [6] [7] [8] [9] [10].

Individual pulses are specified by parameters such as amplitude, duration, rise and fall times. Numerous publications deal with ways to increase pulse voltage or current amplitudes by connecting multiple avalanche transistors in series [2] [5], parallel [1] [5], or in configurations such as the Marx bank [2] [4] [8]. Rise and fall times are largely determined by intrinsic transistor parameters [7] and circuit parasitics [6].

The pulse *repetition rate* is another property of generators that plays an important role in some applications. High repetition rates of ≥20 MHz have been reported for circuits using vacuum tubes [3] and step recovery diodes [11]. One limiting factor in avalanche transistor circuits based on the basic triggered-switched capacitance-discharge topology [1] is the rate at which the energy storage element (lumped or distributed capacitance) at the collector can be recharged. Higher rates are achievable by replacing the charging resistor with a more complex passive [9] [12] or active circuit [10]. Eventually, another limitation arises due to power dissipation in the transistors. It can be mitigated to some extent by heat-sinking, but the pulse rates are still practically limited to a few MHz in continuous operation [10].

The circuit presented in this work (German patent application 10 2017 125 386.6 filed on 10/30/2017) (FIG 1) overcomes both the recharging and dissipation limits. It is based on multiple avalanche transistor sections connected in parallel to the same load and triggered sequentially. A similar principle for increasing the pulse repetition rate of thyratron-based generators has been demonstrated [13], but without the provisions for synchronization detailed in our work.

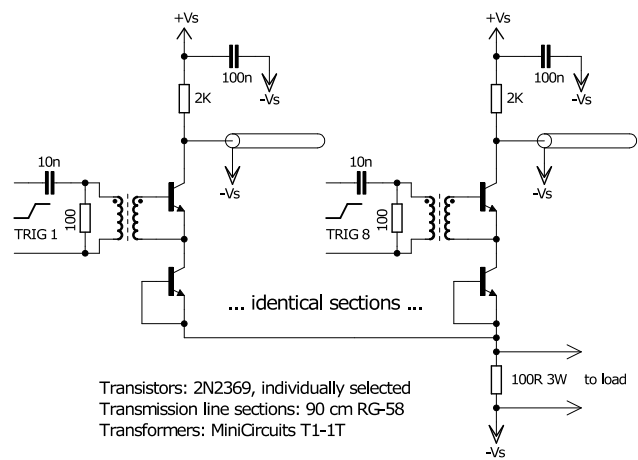

FIG 1. Circuit diagram of the pulse generator circuit

The pulse generator is designed for a specific application: driving the photocathode of a fast-gating image intensifier [14] for synchronized imaging of the ion micromotion in an rf trap [15] at a frequency of about $f$=25 MHz. Micromotion is a generally unwanted driven oscillation which occurs whenever a trapped ion is displaced from the rf node. We determine its amplitude via the photon-correlation technique, which makes use of the fluorescence modulation due to the 1st-order Doppler effect on a strongly allowed optical transition [16], [17].

---

[a] Author to whom correspondence should be addressed: nikolai.beev@cern.ch. Tel: +41227675179. Present address: CERN, CH-1211 Geneva 23, Switzerland.
[b] Present address: National Institute of Standards and Technology, Time and Frequency Division, 325 Broadway, Boulder, CO 80305, USA

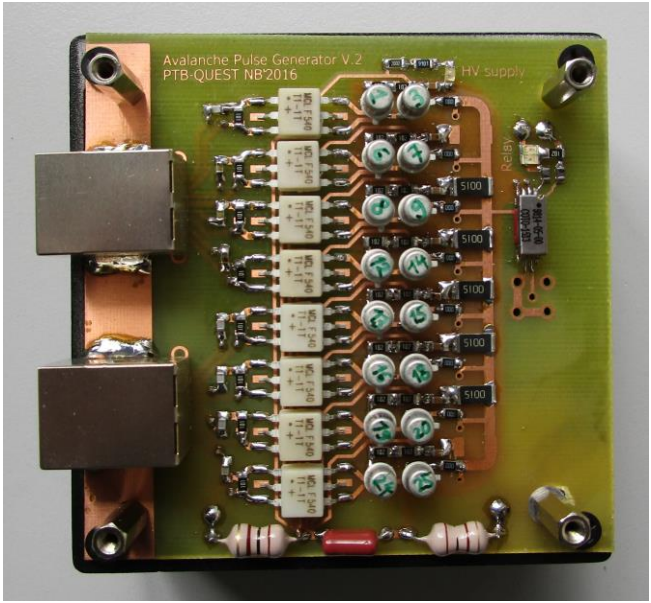

**FIG 2.** Photograph of the 8-section pulse generator circuit

The collected fluorescence at $f$ is demodulated by gating the image intensifier for a fraction of the rf period, with varied delay. We take 4 samples per period, as a compromise between the number of detected photons and the reduction in contrast due to temporal averaging within the gate time. Due to the low number of photons ($4 \times 10^{-5}$ within the 10 ns gate time), a repetition rate of $f$ is desirable to minimize integration times. Photon shot noise is the major uncertainty contribution, and averaging times on the order of about 2 h are necessary to resolve micromotion amplitudes to within 25 pm [18]. In our tests, we operated the circuit for >10 h at a time.

From initial tests with single transistors we determined that a maximum repetition rate of ca. 4 MHz per section is achievable with forced-air cooling. The realized pulse generator (FIG 2) has 8 identical sections. The charge-storage transmission lines are placed on the bottom side of the circuit board, within an enclosure. A strong cooling fan is mounted directly above the shown board. We chose not to heat-sink the transistors, in order to reduce parasitic capacitance added at their case-connected collectors. The cooling fan is essential, as the circuit operated continuously at full pulse rate dissipates about 20 W.

We use the widely available transistor 2N2369, which is known for its good operation in avalanche mode [6]. In order to achieve the same pulse amplitudes from the different sections while having a single supply voltage, we have manually selected transistors with tightly matching breakdown characteristics (threshold voltages matched to ±0.4 V or 1 %). Voltage amplitudes of several tens of V were achieved (FIG 3) with the 2-transistor stack (FIG 1), powered with total supply voltage of 100 - 110 V. These pulse amplitudes are sufficient for gating the image intensifier without appreciable loss of spatial resolution.

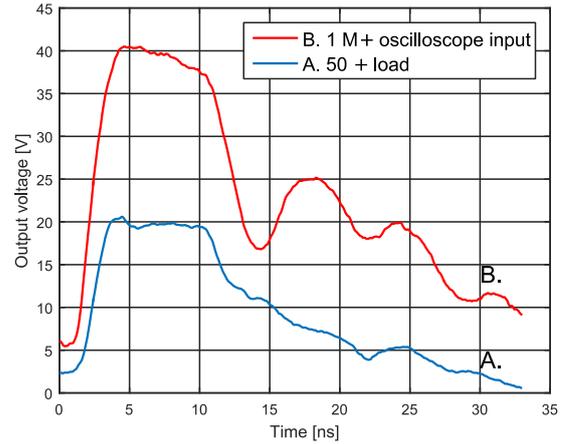

**FIG 3.** Pulse waveforms measured into a 50 Ω load and the high-impedance (1 MΩ, 16 pF) input of a 4 GSPS sampling oscilloscope during continuous operation.

A degradation of pulse edges occurs due to reflections when a reactive load is connected at the output of the generator. The effect is particularly evident in the falling edges of pulses shown in FIG 3. The photocathode of the image intensifier can be seen as a lossy lumped capacitance (C ≈ 100 pF) at the end of a 100 mm long RG-174 coaxial line, which imposes limits on the overall system performance. The achievable temporal resolution is optimized by applying a +15V reverse bias to the cathode, which repels slow photoelectrons generated during the transients and optimally exploits the nonlinear dependence of the device gain on the cathode voltage. FIG 4 shows a calibration measurement of the contrast at $f$=24.4 MHz, normalized to the dc signal. The observed signal is generated by an ion undergoing micromotion, the amplitude of which was determined before and after with a photomultiplier tube (PMT). As the image intensifier measurements consist of temporal averages over 1/4 of the signal period, a contrast reduction by a factor of 0.9 is expected with respect to the PMT signal. No further reduction is seen in the data, confirming that the distortion of the applied pulses has no adverse effect on the observed signal.

The photocathode is gated by application of negative voltage pulses. The image intensifier we use has built-in high-voltage converters for biasing the micro-channel plate (MCP) and the photocathode when used in continuous (non-gated) mode, which permit powering the device from a non-grounded low-voltage supply. The output of our circuit is crossed to invert the polarity of the pulses. A coaxial rf relay allows for selecting between the internal dc photocathode bias and external gating by the pulse generator.

We use a Xilinx Spartan-3 XC3S500E-4FG320C FPGA for generating the synchronized triggering pulses. The rf signal is buffered by a clock-conditioning IC (AD9513) and fed to a global clock input of the FPGA. A secondary clock signal is generated internally using the built-in digital clock manager (DCM) blocks [19] of the FPGA. This secondary clock has an adjustable phase in coarse steps of $90^0$ and $180^0$, controllable by digital inputs. Fine adjustment of the phase is also implemented by tuning the delay-locked loop within the DCM. The secondary clock is used to increment a simple ripple counter that produces an 8-phase sequence of trigger pulses (bottom traces in FIG 5). The outputs of the FPGA are buffered by a CMOS IC (74HC244) with adjustable positive supply voltage. This circuit drives two equally long Cat-5 Ethernet cables, terminated with their characteristic impedance of 100 Ω at the avalanche pulse section (FIG 1).

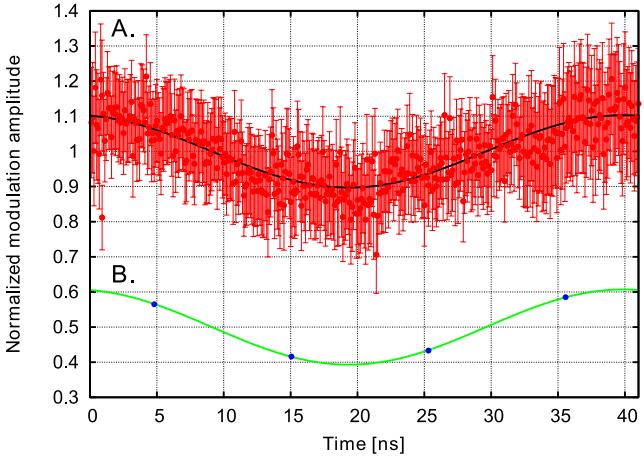

**FIG 4.** Calibration of the contrast at $f$ = 24.4 MHz using modulated fluorescence from an ion undergoing micromotion. **A:** reference measurement using a photomultiplier tube. **B:** CCD camera measurement using the gated image intensifier with an integration time of about 3300 s. The image intensifier data has been vertically offset by -0.5 and a technical phase offset has been subtracted for clarity. The statistical uncertainty of the data points in B. is not shown, as it is smaller than the plot symbols.

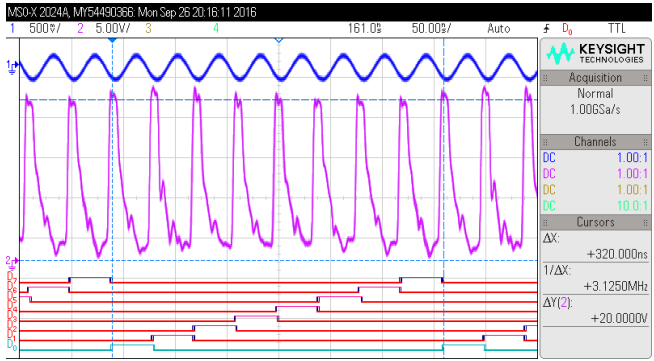

**FIG 5.** Oscilloscope screenshot showing the trigger signals D0-D7 (bottom traces), pulse generator output into 50 Ω (middle trace) and 25 MHz rf signal (top trace)

In summary, we have devised a way to increase the pulse repetition rate of a generator built of multiple parallel avalanche transistor sections, and have demonstrated its operation at 25 MHz with a fast-gating image intensifier. It is straightforward to extend the presented principle to even higher repetition rates by increasing the number of sections (cf. FIG. 1). The only practical limitations would arise from the increasing circuit size, parasitics and total power dissipation.

We acknowledge financial support from DFG through CRC 1227 (DQ-mat), project B03, and grant ME 3648/1-1. We thank Michael Drewsen of Aarhus University for providing a reference design for a MOSFET-based pulse generator. We are also grateful to Dirk Piester from PTB, Department 4.42, for lending us a fast sampling oscilloscope.